\documentclass[preprint, prX]{revtex4}

\usepackage{amsmath}    
\usepackage{graphicx}   
\usepackage{verbatim}   
\usepackage{color}      
\usepackage{subfigure}  
\usepackage{hyperref}   
\usepackage{amssymb}    
\usepackage{epsfig}
\usepackage{graphics,graphicx}
\usepackage{setspace}
\usepackage{url}
\usepackage{algorithm,algorithmic}
\usepackage{MnSymbol}

\begin{document}

\begin{abstract}
A manual measuring time tool in mass sporting competitions would not be imaginable nowadays, because many modern disciplines, such as
IRONMAN, last a long-time and, therefore, demand additional reliability. Moreover, automatic timing-devices based on RFID technology,
have become cheaper. However, these devices cannot operate as stand-alone because they need a computer measuring system that is capable of processing incoming events, encoding the results, assigning them to the correct competitor, sorting the results according to the
achieved times, and then providing a printout of the results. This article presents the domain-specific language EasyTime, which enables the controlling of an agent by writing the events within a database. It focuses, in particular, on the implementation of EasyTime with a LISA tool that enables the automatic construction of compilers from language specifications, using Attribute Grammars.

\textit{To cite paper as follows: I. Jr. Fister, M. Mernik, I. Fister, D. Hrn\v{c}i\v{c}. 
Implementation of EasyTime Formal Semantics using a LISA Compiler Generator. Computer Science
and Information Systems, 2012, in press.
}

\end{abstract}

\title{Implementation of EasyTime Formal Semantics using a LISA Compiler Generator}

\author{Iztok Fister Jr.}
\altaffiliation{University of Maribor, Faculty of electrical engineering and computer science
Smetanova 17, 2000 Maribor}
\email{iztok.fister@guest.arnes.si}

\author{Marjan Mernik}
\altaffiliation{University of Maribor, Faculty of electrical engineering and computer science
Smetanova 17, 2000 Maribor}
\email{marjan.mernik@uni-mb.si}

\author{Iztok Fister}
\altaffiliation{University of Maribor, Faculty of electrical engineering and computer science
Smetanova 17, 2000 Maribor}
\email{iztok.fister@uni-mb.si}

\author{Dejan Hrn\v{c}i\v{c}}
\altaffiliation{University of Maribor, Faculty of electrical engineering and computer science
Smetanova 17, 2000 Maribor}
\email{dejan.hrncic@uni-mb.si}

\maketitle

\section{Introduction}
In the past, timekeepers measured the time manually. The time given by a timer was assigned to competitors based on their starting number, and these competitors were then placed in order according to their achieved results and category. Later, manual timers were replaced by timers with automatic time-registers capable of capturing and printing out registered times. However, assigning the times to competitors based on their starting numbers, was still done manually. This work could be avoided by using electronic-measuring technology which, in addition to registering the time, also enables the registering of competitors' starting numbers. An expansion of RFID (Radio Frequency Identification) technology has helped this measuring-technology to become less expensive (\cite{web:ChampionChip2010,web:RFID2010}), and accessible to a wider-range of users (e.g., sports clubs, organizers of sporting competitions). Moreover, they were also able to compete with time-measuring monopolies at smaller competitions.

In addition to measuring technology, a flexible computer system is also needed to monitor the results. The proposed computer system enables the monitoring of different sporting competitions using a various number of measuring devices and measuring points, the online recording of events, the writing of results, as well as efficiency and security. This measuring device is dedicated to the registration of events and is triggered either automatically, when the competitor crosses the measuring point that acts as an electromagnetic antenna fields with an appropriate RFID tag, or manually, when an operator presses the suitable button on a personal computer that acts as a timer. The control point is the place where the organizers want to monitor the results. Until now, each control point has required its own measuring device. However, modern electronic-measuring devices now allow for the handling of multiple control points, simultaneously. Moreover, each registered event can have a different meaning, depending on the situation within which it is generated. Therefore, an event is handled by the measuring system according to those rules that are valid for the control point. As a result, the number of control points (and measuring devices) can be reduced by using more complex measurements. Fortunately, the rules controlling events can be described easily with the use of a domain-specific language (DSL) \cite{Hudak:1996,Mernik:2005}. When using this DSL, measurements at different sporting competitions can be accomplished by an easy pre-configuration of the rules.

A DSL is suited to an application domain and has certain advantages over general-purpose languages (GPL) within a specific domain
\cite{Mernik:2005}. The GPL is dedicated to writing software over a wider-range of application domains. General problems are usually solved using these languages. However, a programmer is necessary for changing the behavior of a program written in a GPL. On the
other hand, the advantages of DSL are reflected in its greater expressive power in a particular domain and, hence, increased productivity~\cite{Kosar:ESE:2011} , ease of use (even for those domain experts who are not programmers), and easier verification and optimization \cite{Mernik:2005}. This article presents a DSL called EasyTime, and its implementation. EasyTime is intended for controlling those agents responsible for recording events from the measuring devices, into a database. Therefore, the agents are crucial elements of the proposed measuring system. To the best of the author's knowledge there is no comparable DSL of time measuring for sport events, whilst some DSLs for performance measurement of computer systems ~\cite{Arpaia:2011,Pakin:2007} as well as on general measurement systems do indeed already exist~\cite{Kos:2011}. Finally, EasyTime has been successfully employed in practice, as well. For instance, it measured times at the World Championship for the double ultra triathlon in 2009~\cite{Fister:2011}, and at a National Championship in the time-trials for bicycle in 2010~\cite{Fister:2011}.

The structure of the remaining article is as follows; In the second section, those problems are illustrated that accompany time-measuring at sporting competitions. Focus is directed primarily on triathlon competitions, because they contain three disciplines that need to be measured, and also because of their lengthy durations. The design of DSL EasyTime is briefly shown in section three. The implementation of the EasyTime compiler is described in the fourth section, whilst the fifth section explains the execution of the program written in EasyTime. Finally, the article is concluded with a short analysis of the work performed, and a look at future work.
This paper extends a previous workshop paper \cite{Fister:2011a} by providing general guidelines on how to transform formal language specifications using denotational semantics into attribute grammars. The concreteness of these guidelines is shown on EasyTime DSL.

\section{Measuring Time in Sporting Competitions}

In practice, the measuring time in sporting competitions can be performed manually (classically or with a computer timer) or automatically (with a measuring device). The computer timer is a program that usually runs on a workstation (personal computer) and measures in real-time. Thereby, the processor tact is exploited. The processor tact is the velocity with which the processor's instructions are interpreted. A computer timer enables the recording of events that are generated by the competitor crossing those measure points (MP) in line with the measuring device. In that case, however, the event is triggered by an operator pressing the appropriate button on the computer. The operator generates events in the form of $\langle\#,MP,TIME\rangle$, where $\#$ denotes the starting number of a competitor, $MP$ is the measuring point, and $TIME$ is the number of seconds since 1.1.1970 at 0:0:0 (timestamp). One computer timer represents one measuring-point.

Today, the measuring device is usually based on RFID technology \cite{Finkenzeller:2010}, where
identification is performed using electromagnetic waves within a range of radio frequencies, and consists of the following elements:
\begin{itemize}
  \item readers of RFID tags,
  \item primary memory,
  \item LCD monitor,
  \item numerical keyboard, and
  \item antenna fields.
\end{itemize}
More antenna fields can be connected on to the measuring device. One antenna field represents one measuring point. Each competitor
generates an event by crossing the antenna field using passive RFID tags that include an identification number. This number is unique and differs from the starting number of the competitor. The event from the measuring device is represented in the form of
$\langle\#,RFID,MP,TIME\rangle$, where the identification number of the RFID tag is added to the previously mentioned triplet.

The measuring devices and workstations running the computer timer can be connected to the local area network. Communication with devices is performed by a monitoring program, i.e. an agent, that runs on the database server. This agent communicates with the measuring device via the TCP/IP sockets, and appropriate protocol. Usually, the measuring devices support a $Telnet$ protocol that is character-stream
oriented and, therefore, easy to implement. The agent employs the file transfer protocol ($ftp$) to communicate with the computer timer.

\subsection{Example: Measuring Time in Triathlons}

Special conditions apply for triathlon competitions, where one competition consists of three disciplines. This article,
therefore, devotes most of its attention to this problem.

The triathlon competition is performed as follows: first, the athletes swim, then they ride a bicycle and finally run. In practice,
all these activities are performed consecutively. However, the transition times, i.e. the time that elapses when a competitor shifts
from swimming to bicycling, and from bicycling to running, are added to the summary result. There are various types of triathlon
competitions that differ according to the lengths of various courses. In order to make things easier, organizers often employ round
courses (laps) of shorter lengths instead of one long course. Therefore, the difficulty of measuring time is increased because the time
for each lap needs to be measured.

\begin{figure*}[htb]  
\vspace{-5mm}
    \begin{center}
        \includegraphics [scale=0.9]{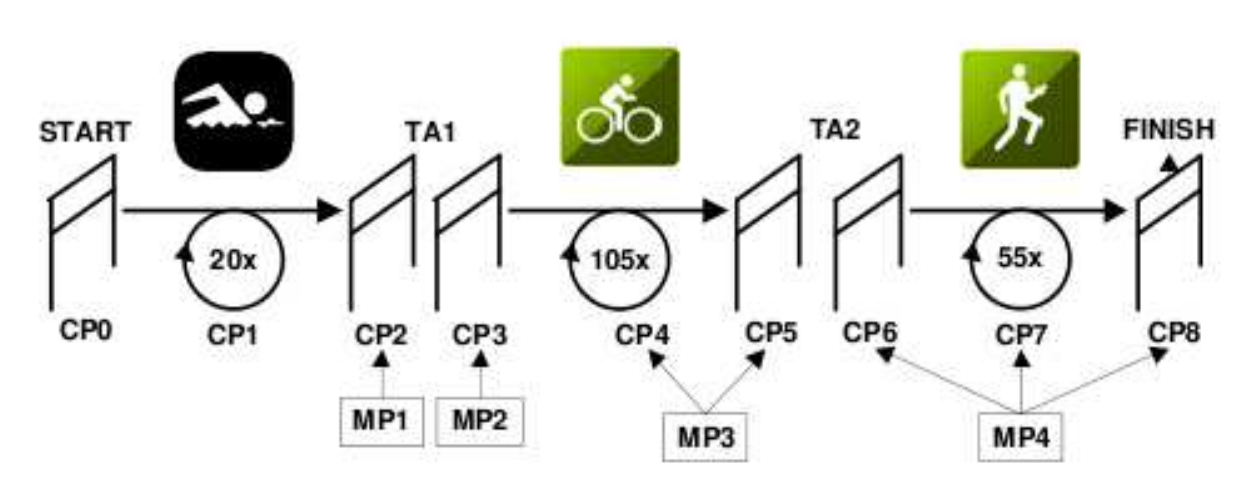}  %
        \caption{Definition of control points in the triathlon}
        \label{pic:slika_1}
    \end{center}
\vspace{-5mm}
\end{figure*}

Measuring time in triathlon competitions can be divided into nine control points (Fig.~\ref{pic:slika_1}). The control point (CP) is a
location on the triathlon course, where the organizers need to check the measured time. This can be intermediate or final. When dealing
with a double triathlon there are 7.6 km of swimming, 360 km of bicycling, and 84 km of running. Hence the swimming course of 380 meters
consists of 20 laps, the bicycling course of 3.4 kilometers contains 105 laps, and the running course of 1.5 kilometers has 55
laps (Fig.~\ref{pic:slika_1}).

Therefore, the final result for each competitor in a triathlon competition (CP8) consists of five final results: the swimming time SWIM
(CP2-CP0), the time for the first transition TA1 (CP3-CP2), the time spent bicycling BIKE (CP5-CP3), the time for the second transition TA2 (CP6-CP5), the time spent running RUN (CP8-CP6), and three intermediate results: the intermediate time for swimming (CP1), the intermediate time for bicycling (CP4) and the intermediate time for running (CP7). However, the current time INTER\_x and the number of remaining laps LAPS\_x are measured by the intermediate results, where $x=\{1,2,3\}$ denotes the appropriate discipline (1=SWIM, 2=BIKE and 3=RUN).

The DSL EasyTime was developed in order to achieve this goal, and has been employed in practice by conducting measurements at the World Championship in the Double Triathlon in 2009. Note that the measurements were realized according to Fig.~\ref{pic:slika_1}. The next sections presents the design, implementation, and operation of EasyTime.

\section{The Design of the EasyTime Domain-Specific Language}

Typically, the development of a DSL consists of the following phases~\cite{Mernik:2005}:
\begin{itemize}
  \item a domain analysis,
  \item a definition of an abstract syntax,
  \item a definition of a concrete syntax,
  \item a definition of formal semantics, and
  \item an implementation of the DSL.
\end{itemize}

Domain analysis provides an analysis of the application domain, i.e. measuring time in sporting competitions. The results of this
analysis define those concepts of EasyTime that are typically represented within a feature diagram~\cite{Deursen:2002,Stuikys:2009}. The feature diagram also describes dependencies between the concepts of DSL. Thus, each concept can be broken-down into features and sub-features. In the case of EasyTime, the concept $race$ consists of sub-features: $events$ (e.g., $swimming$, $bicycling$, and $running$), $control\ points$, $measuring\ time$, $transition$ $area$, and $agents$. Each $control\ point$ is described by its $starting$ and $finish$ line and at least one $lap$. In addition, the feature $transition\ area$ can be introduced as the difference between the finish and start times. Both $updating\ time$ and $decrementing\ laps$ are sub-features of $measuring\ time$. However, an $agent$ is needed for the processing of events received from the measuring device. It can act either $automatically$ or $manually$.
Note that during domain analysis not all the identified concepts are useful for solving actual problem. Hence, the identified concepts can be further classified into ~\cite{Mauw:2004}:

\begin{itemize}
  \item irrelevant concepts, those which are irrelevant to the actual problem;
  \item variable concepts, those which actually need to be described in the DSL program; and
  \item fixed concepts, those which can be built into the DSL execution environment.
\end{itemize}

Domain analysis identifies several variable and fixed concepts within the application domain that needs to be mapped into EasyTime syntax and semantics~\cite{Mernik:2005}. At first, the abstract syntax is defined (context-free grammar). Each variable concept obtained from the domain
analysis is mapped to a non-terminal in the context-free grammar; additionally, some new non-terminal and terminal symbols are defined. The translations of the EasyTime domain concepts to non-terminals are presented and explained in Table~\ref{tab:tab1}, whilst an abstract syntax is presented in Table~\ref{tab:X}.
Note that, the concepts \textit{Events} and \textit{Transition} are irrelevant for solving actual problem and are not mapped into non-terminals' symbols (denoted as \textit{none} in Table~\ref{tab:tab1}).
Interestingly, a description of agents and measuring places cannot be found in other DSLs or GPLs.
Whilst attribute declaration is similar to variable declaration in many other programming languages.
However, note that there is the distinction that variables are actually database attributes allocated for every competitor. Some statements, such as assignment, conditional statement, and compound statement can be found in many other programming languages, whilst decrement attributes and update attributes are domain-specific constructs.

\begin{table*}[htb]
\caption{Translation of the application domain concepts into a context-free grammar}
\label{tab:tab1}
\vspace{-5mm}
\scriptsize
\begin{center}
\begin{tabular}{ l  l  l  l }
\hline
Application domain concepts &  Non-terminal & Formal sem. & Description \\

\hline
Race & P & $\mathcal{CP}$ & Description of agents; control points; measuring \\
 & & & places. \\
\hline
Events (swimming, cycling, & none & none & Measuring time is independent from the type of an \\
running) & & & event. However, good attribute's identifier in control \\
 & & & points description will resemble the type of an event. \\
\hline
Transition area times & none & none & Can be computed as difference between events final \\
 & & & and starting times. \\
\hline
Control points (start, number & D & $\mathcal{D}$ & Description of attributes where start and finish time \\
of laps, finish) & & & will be stored as well as remaining laps. \\
\hline
Measuring places (update time, & M & $\mathcal{CM}$ & Measuring place id; agent id, which will control this \\
decrement lap) & & & measuring place; specific actions (presented \\
 & & & with new non-terminal S) which will be performed \\
 & & & at this measuring place (e.g., decrement lap). \\
\hline
Agents (automatic, manual) & A & $\mathcal{A}$ & Agent id; agent type (automatic, manual); agent sour- \\
 & & & ce (file, ip). \\
\hline
\end{tabular}
\end{center}
\normalsize
\vspace{-5mm}
\end{table*}

\begin{table}[htb]           
\caption{The abstract syntax of EasyTime}
\label{tab:X}
\vspace{-5mm}
\footnotesize
\begin{center}
\begin{tabular}{ | l  l  l | }
\hline
  $P \in $ \textbf{Pgm} & &  $A \in $ \textbf{Adec} \\
  $D \in $ \textbf{Dec} & &  $M \in $ \textbf{MeasPlace} \\
  $S \in $ \textbf{Stm} & &  $b \in $ \textbf{Bexp} \\
  $a \in $ \textbf{Aexp} & &  $n \in $ \textbf{Num} \\
  $x \in $ \textbf{Var} & &  $file \in $ \textbf{FileSpec} \\
  $ip \in $ \textbf{IpAddress} & &  \\
  & & \\
  $P$ & ::= & $A\ D\ M$ \\
  $A$ & ::= & $n$ \textbf{manual} $file$ \textbar $\ n$ \textbf{auto} $ip$ \textbar $\ A_{1};A_{2}$ \\
  $D$ & ::= & \textbf{var} $x := a$ \textbar $\ D_{1};D_{2}$ \\
  $M$ & ::= & \textbf{mp}[$n_{1}$] $\rightarrow$ \textbf{agnt}[$n_{2}]\ S$ \textbar $\ M_{1};M_{2}$ \\
  $S$ & ::= & \textbf{dec} $x$ \textbar \ \textbf{upd} $x$ \textbar $\ x := a$ \textbar $\ (b) \rightarrow S$ \textbar $\ S_{1};S_{2}$
  \\
  $b$ & ::= & \textbf{true} \textbar \ \textbf{false} \textbar $\ a_{1} == a_{2}$ \textbar $\ a_{1} != a_{2}$ \\
  $a$ & ::= & $n$ \textbar $\ x$ \\
\hline
\end{tabular}
\end{center}
\normalsize
\vspace{-5mm}
\end{table}

Although a language designer can proceed after domain analysis with informal or formal design patterns~\cite{Mernik:2005} the formal design step is preferred since it can identify problems before the DSL is actually implemented~\cite{Viroli:2011}.
Moreover, formal specifications can be implemented automatically by language development systems, thus
significantly reducing the implementation effort~\cite{Mernik:2005}.
The meaning of the EasyTime language constructs is prescribed during the formal semantics phase. Each language construct, belonging to the syntax domain, is mapped into an appropriate semantic domain (Table~\ref{tab:Y}) by semantic functions $\mathcal{CP}$, $\mathcal{A}$, $\mathcal{D}$, $\mathcal{CM}$, $\mathcal{CS}$,
$\mathcal{CB}$, and $\mathcal{CA}$ (Table~\ref{tab:tab6}).

\begin{table}[htb]        
\caption{Semantic domains}
\label{tab:Y}
\footnotesize
\vspace{-5mm}
\begin{center}
\begin{tabular}{ | l  l | }
\hline
  \textbf{Integer}=$\{\ldots -3,-2,-1,0,1,2,3 \ldots\}$ & $n \in$ \textbf{Integer} \\
  \textbf{Truth-Value}=$\{true,false\}$ &  \\
  \textbf{State}=\textbf{Var}$\rightarrow$\textbf{Integer} & $s \in$ \textbf{State} \\
  \textbf{AType}=$\{manual,auto\}$ &  \\
  \textbf{Agents}=\textbf{Integer}$\rightarrow$\textbf{AType}$\ \times\ (FileSpec\ \cup\ IpAddress)$ & $ag \in$\textbf{Agents} \\
  \textbf{Runners}=$(Id \times RFID \times LastName \times FirstName)^{*}$ & $r \in $ \textbf{Runners}\\
  \textbf{DataBase}=$(Id \times Var_{1} \times Var_{2} \times \ldots \times Var_{n})^{*}$ & $db \in $ \textbf{DataBase}\\
  \textbf{Code}=\textbf{String} & $c \in $ \textbf{Code} \\
\hline
\end{tabular}
\end{center}
\normalsize
\vspace{-5mm}
\end{table}

\begin{table}[!htb]
\caption{EasyTime formal semantics}
\label{tab:tab6}
\footnotesize
\vspace{-5mm}
\begin{center}
\begin{tabular}{ | l  l  l | }
\hline
  $\mathcal{CP}:\textbf{Pgm} \rightarrow \textbf{Runners}$ & $\rightarrow$ & \textbf{Code} $\times$ \textbf{Integer} $\times$ \textbf{DataBase} \\
  $\mathcal{CP} \lsem A\ D\ M \rsem r$ & = & let $s= \mathcal{D} \lsem D \rsem \O$: \\
  & & $\>\>\>db = $\textnormal{create\&insertDB}$(s,r)$ \\
  & & in $(\mathcal{CM}\lsem M \rsem (\mathcal{A}\lsem A \rsem \O), db)$ \\
  & & \\
  $\mathcal{A}$~:~\textbf{Adec} $\rightarrow$ \textbf{Agents} & $\rightarrow$ & \textbf{Agents} \\
  $\mathcal{A} \lsem n$ \textbf{manual} $ file\rsem ag$ & = & $ag [ n \rightarrow (manual, file) ]$ \\
  $\mathcal{A} \lsem n$ \textbf{auto} $ ip\rsem ag$ & = & $ag [ n \rightarrow (auto, ip) ]$ \\
  $\mathcal{A} \lsem A_{1};A_{2}\rsem ag$ & = & $\mathcal{A} \lsem A_{2} \rsem (\mathcal{A} \lsem A_{1} \rsem ag)$ \\
  & & \\
  $\mathcal{D}$~:~\textbf{Dec}$\rightarrow$\textbf{State} & $\rightarrow$ & \textbf{State} \\
  $\mathcal{D} \lsem \textbf{var}\ x := a \rsem s$ & = & $s[x \rightarrow a]$ \\
  $\mathcal{D} \lsem D_{1},D_{2} \rsem s$ & = & $\mathcal{D} \lsem D_{2} \rsem (\mathcal{D} \lsem D_{1} \rsem s)$\\
  & & \\
  $\mathcal{CM}$~:~\textbf{MeasPlace} $\rightarrow$ \textbf{Agents}&$\rightarrow$&\textbf{Code} $\times$ \textbf{Integer} \\
  $\mathcal{CM} \lsem \textbf{mp}[n_{1}] \rightarrow \textbf{agnt}[n_{2}] S \rsem ag$&=&(WAIT $i:\mathcal{CS} \lsem S \rsem (ag, n_{2}),
  n_{1} )$  \\
  $\mathcal{CM} \lsem M_{1}; M_{2} \rsem ag$&=&$\mathcal{CM} \lsem M_{1} \rsem ag: \mathcal{CM} \lsem M_{2} \rsem ag$ \\
  & & \\
  $\mathcal{CS}$~:~\textbf{Stm}$\rightarrow$ \textbf{Agents} $\times$ \textbf{Integer} & $\rightarrow$ & \textbf{Code} \\
  $\mathcal{CS} \lsem$ \textbf{dec} $x \rsem (ag,n)$ & = & FETCH $x$:DEC:STORE $x$ \\
  $\mathcal{CS} \lsem$ \textbf{upd} $x \rsem (ag,n)$ & = & FETCH $y$:STORE $x\ $ where \\ & & $y = \left\{\begin{array}{l l}
  \textnormal{accessfile}(ag(n)\downarrow 2) & \textnormal{if}\ ag(n)\downarrow 1 = manual \\
  \textnormal{connect}(ag(n)\downarrow 2) & \textnormal{if}\ ag(n)\downarrow 1 = automatic \\
  \end{array}\right.$ \\
  $\mathcal{CS} \lsem x := a \rsem(ag,n)$ & = & $\mathcal{CA}\lsem a\rsem$:STORE $x$ \\
  $\mathcal{CS} \lsem (b)\rightarrow S \rsem(ag,n)$ & = & $\mathcal{CB} \lsem b\rsem$:BRANCH($\mathcal{CS}\lsem S\rsem (ag,n),NOOP$)\\ $\mathcal{CS} \lsem S_{1};S_{2} \rsem(ag,n)$ & = &
  $\mathcal{CS}\lsem S_{1}\rsem(ag,n):\mathcal{CS}\lsem S_{2}\rsem (ag,n)$ \\
  & & \\
  $\mathcal{CB}$~:~\textbf{Bexp} & $\rightarrow$ & \textbf{Code} \\
  $\mathcal{CB} \lsem \textbf{true} \rsem $ & = & TRUE \\
  $\mathcal{CB} \lsem \textbf{false} \rsem $ & = & FALSE \\
  $\mathcal{CB} \lsem a_{1}==a_{2} \rsem$ & = & $\mathcal{CA} \lsem a_{2}\rsem:\mathcal{CA} \lsem a_{1}\rsem$:EQ \\
  $\mathcal{CB} \lsem a_{1}!=a_{2} \rsem$ & = & $\mathcal{CA} \lsem a_{2}\rsem:\mathcal{CA} \lsem a_{1}\rsem$:NEQ \\
  & & \\
  $\mathcal{CA}$~:~\textbf{Aexp} & $\rightarrow$ & \textbf{Code} \\
  $\mathcal{CA} \lsem n \rsem$ & = & PUSH $n$ \\
  $\mathcal{CA}\lsem x \rsem$ & = & FETCH $x$ \\
\hline
\end{tabular}
\end{center}
\normalsize
\vspace{-5mm}
\end{table}

\begin{algorithm}[htb]
\caption{EasyTime program for measuring time in a triathlon competition as illustrated in Fig.~\ref{pic:slika_1}}
\label{alg:prog}
\scriptsize
\begin{algorithmic}[1]
\STATE 1 manual "abc.res";
\STATE 2 auto 192.168.225.100;
\STATE
\STATE var ROUND1 := 20;
\STATE var INTER1 := 0;
\STATE var SWIM := 0;
\STATE var TRANS1 :=0;
\STATE var ROUND2 := 105;
\STATE var INTER2 :=0;
\STATE var BIKE := 0;
\STATE var TRANS2 :=0;
\STATE var ROUND3 := 55;
\STATE var INTER3 := 0;
\STATE var RUN := 0;
\STATE
\STATE mp[1] $\rightarrow$ agnt[1] \{
\STATE \ \ (true) $\rightarrow$ upd SWIM;
\STATE \ \ (true) $\rightarrow$ dec ROUND1;
\STATE \}
\STATE mp[2] $\rightarrow$ agnt[1] \{
\STATE \ \ (true) $\rightarrow$  upd TRANS1;
\STATE \}
\STATE mp[3] $\rightarrow$  agnt[2] \{
\STATE \ \ (true) $\rightarrow$  upd INTER2;
\STATE \ \ (true) $\rightarrow$  dec ROUND2;
\STATE \ \ (ROUND2 == 0) $\rightarrow$  upd BIKE;
\STATE \}
\STATE mp[4] $\rightarrow$  agnt[2] \{
\STATE \ \ (ROUND3 == 55) $\rightarrow$  upd TRANS2;
\STATE \ \ (true) $\rightarrow$  upd INTER3;
\STATE \ \ (true) $\rightarrow$  dec ROUND3;
\STATE \ \ (ROUND3 == 0) $\rightarrow$  upd RUN;
\STATE \}
\end{algorithmic}
\normalsize
\end{algorithm}

These semantic functions translate EasyTime constructs into the instructions of the simple virtual machine. The meaning of virtual machine instructions has been formally defined using operational semantics (Table~\ref{tab:am}) as the transition of configurations $<c,~e,~db,~j>$, where $c$ is a sequence of instructions, $e$ is the evaluation stack to evaluate arithmetic and boolean expressions, $db$ is the database, and $j$ is the starting number of a competitor. More details of EasyTime syntax and semantics are presented in~\cite{Fister:2011}. This article focuses on the implementation phase, as presented in the next section.

\begin{table*}[htb]       
\caption{The virtual machine specification}
\label{tab:am}
\begin{center}
\vspace{-5mm}
\scriptsize
\begin{tabular}{ | l  l  l  l | }
\hline
  $\langle \textnormal{PUSH}\ n:c,e,db,j \rangle$ & $\triangleright$ & $\langle c,n:e,db,j \rangle$ & \\
  $\langle \textnormal{TRUE}:c,e,db,j \rangle$ & $\triangleright$ & $\langle c,true:e,db,j \rangle$ & \\
  $\langle \textnormal{FALSE}:c,e,db,j \rangle$ & $\triangleright$ & $\langle c,false:e,db,j \rangle$ & \\
  $\langle \textnormal{EQ}:c,z_{1}:z_{2}:e,db,j \rangle$ & $\triangleright$ & $\langle c,(z_{1}==z_{2}):e,db,j \rangle$ & \textnormal{\ if}\ $z_{1},z_{2} \in \textbf{Int}$ \\
  $\langle \textnormal{NEQ}:c,z_{1}:z_{2}:e,db,j \rangle$ & $\triangleright$ & $\langle c,(z_{1}!=z_{2}):e,db,j \rangle$ & \textnormal{\ if}\ $z_{1},z_{2} \in \textbf{Int}$ \\
  $\langle \textnormal{DEC}:c,z:e,db,j \rangle$ & $\triangleright$ & $\langle c,(z-1):e,db,j \rangle$ & \textnormal{\ if}\ $z \in \textbf{Int}$ \\
  $\langle \textnormal{WAIT}\ i:c,e,db,j \rangle$ & $\triangleright$ & $\langle c,e,db,i \rangle$ & \\
  $\langle \textnormal{FETCH}\ x:c,e,db,j \rangle$ & $\triangleright$ & $\langle c,\textnormal{select}\ x\ \textnormal{from}\ db\ \textnormal{where}\ Id=j:e,db,j \rangle$ & \\
  $\langle \textnormal{FETCH}\ accessfile(fn):c,e,db,j \rangle$ & $\triangleright$ & $\langle c,time:e,db,j \rangle$ & \\
  $\langle \textnormal{FETCH}\ connect(ip):c,e,db,j \rangle$ & $\triangleright$ & $\langle c,time:e,db,j \rangle$ & \\
  $\langle \textnormal{STORE}\ x:c,z:e,db,j \rangle$ & $\triangleright$ & $\langle c,e,\textnormal{update}\ db\ \textnormal{set}\ x=z\ \textnormal{where}\ Id=j,j \rangle$ & \textnormal{\ if}\ $z \in
  \textbf{Int}$ \\
  $\langle \textnormal{NOOP}:c,e,db,j \rangle$ & $\triangleright$ & $\langle c,e,db,j \rangle$ & \\
  $\langle \textnormal{BRANCH}(c_{1},c_{2}):c,t:e,db,j \rangle$ & $\triangleright$ &  $\left\{\begin{matrix}
  \langle c_{1}:c,e,db,j \rangle \\
  \langle c_{2}:c,e,db,j \rangle
  \end{matrix}\right.$ & $\begin{array}{l l} \\
  \textnormal{if}\ t=true \\
  \textnormal{otherwise} \\
  \end{array}$ \\
\hline
\end{tabular}
\normalsize
\vspace{-5mm}
\end{center}
\end{table*}

The sample program written in EasyTime that covers the measuring time in the double ultra triathlon is presented by Algorithm~\ref{alg:prog}. In lines 1-2 two agents are defined. Agent no. 1 is manual and agent no. 2 is automatic. In lines 4-14 several variables, attributes in a database for each competitor, are defined and initialized appropriately. For example, from Figure \ref{pic:slika_1} it can be seen that 20 laps are needed for the swimming course and $ROUND1$ is set to 20, 105 laps are needed for the bicycling course and $ROUND2$ is set to 105, and 55 laps are needed for the running course and $ROUND3$ is set to 55. Lines 16-19 define the first measuring place which is controlled by manual agent no. 1. At this measuring place the intermediate swimming time must be updated in the database ($upd~SWIM$) and the number of laps must be decremented ($dec~ROUND1$). Lines 20-22 define the second measuring place which is also controlled by manual agent no. 1. At this measuring place only transition time must be stored in the database ($upd~TRANS1$). Lines 23-27 define the third measuring place which is controlled by automatic agent no. 2. At this measuring place we must update the intermediate result for bicycling ($upd~INTER2$) and decrement the number of laps ($dec~ROUND2$). If a competitor finished all the requested 105 laps ($ROUND2==0$) then time spent on the bicycle must be stored in the database ($upd~BIKE$). Lines 28-33 define the fourth measuring place which is also controlled by automatic agent no. 2. At this measuring place we must first check if a competitor has just started running ($ROUND3==55$). If this is the case, we must record the transition time between bicycling and running ($upd~TRANS2$). At this measuring place we also must update the intermediate result for running ($upd~INTER3$) and decremented number of laps ($dec~ROUND3$). If a competitor finished all the requested 55 laps ($ROUND3==0$) then the final time must be stored in the database ($upd~RUN$).

\section{Implementation of the Domain-Specific Language EasyTime}

\subsection {A LISA Compiler-Generator}

\begin{figure*}[htb]      
    \begin{center}
        \includegraphics [scale=0.8] {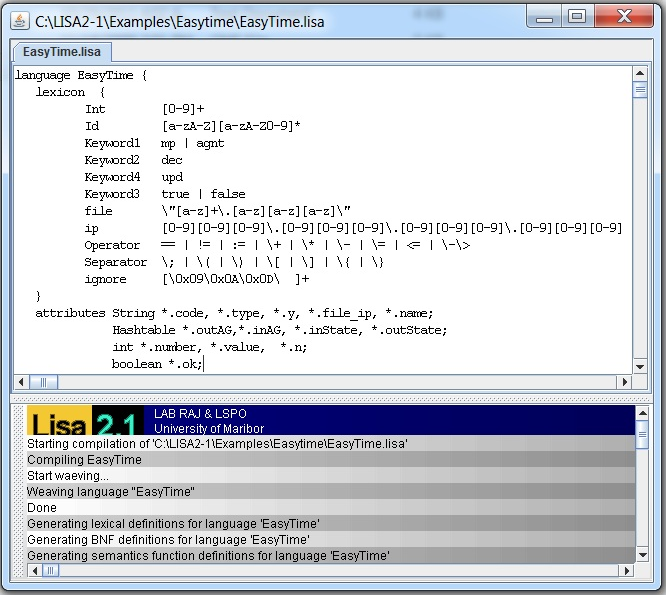}  
        \caption{LISA GUI}
        \label{pic:LISA_GUI}
    \end{center}
\vspace{-5mm}
\end{figure*}

One of the benefits of formal language specifications is the unique possibility for automatic language implementation. Although some compiler generators accept denotational semantics~\cite{Paulson:1982}, the generated compilers are mostly inefficient.
Although many compiler-generators based on attribute grammars~\cite{Knuth:1968,Paakki:1995} exist today, we selected a LISA compiler-compiler that was developed at the University of Maribor in the late 1990s~\cite{Mernik:2002}. The LISA tool produces a highly efficient source code for: the scanner, parser, interpreter or compiler, in Java. The lexical and syntactical parts of the language specification in LISA supports various well-known formal methods, such as regular expressions and BNF~\cite{Aho:1972}. LISA provides two kinds of user interfaces:
\begin{itemize}
  \item a graphic user interface (GUI) (Fig.~\ref{pic:LISA_GUI}), and
  \item a Web-Service user interface.
\end{itemize}

The main features of LISA are as follows:
\begin{itemize}
 \item since it is written in Java, LISA works on all Java platforms,
 \item a textual or a visual environment,
 \item an Integrated Development Environment (IDE), where users can specify, generate, compile and execute programs on the fly,
 \item visual presentations of different structures, such as finite-state-automata, BNF, a dependency graph, a syntax tree, etc.,
 \item modular and incremental language development~\cite{Mernik:2005a}.
\end{itemize}
LISA specifications are based on Attribute Grammar (AG)~\cite{Paakki:1995} as introduced by D.E. Knuth~\cite{Knuth:1968}.
The attribute grammar is a triple $AG=\langle G,A,R \rangle$, where $G$ denotes a context-free grammar, $A$ a finite set of attributes,
and $R$ a finite set of semantic rules. In line with this, the LISA specifications (Table~\ref{tab:tab30}) include:
\begin{itemize}
  \item lexical regular definitions (lexicon part in Table~\ref{tab:tab30}),
  \item attribute definitions (attributes part in Table~\ref{tab:tab30}),
  \item syntax rules (rule part before compute in Table~\ref{tab:tab30}),
  \item semantic rules, (rule part after compute in Table~\ref{tab:tab30}) and
  \item operations on semantic domains (method part in Table~\ref{tab:tab30}).
\end{itemize}

\begin{table}[htb]           
\caption{LISA specifications}
\label{tab:tab30}
\vspace{-5mm}
\footnotesize
\begin{center}
\begin{tabular}{ | l | }
\hline
\ \ {\bf language } $L_{1}$ $[$ {\bf extends} $L_{2}$, ..., $L_{N}$$]$ \{\\
\ \ \ \ {\bf lexicon} \{\\
\ \ \ \ \ \ $[[$P$]$ {\bf overrides} $|$ $[$P$]$ {\bf extends}$]$ R regular expr.\\
\ \ \ \ \ \ \ \ \vdots\\
\ \ \ \ \}\\
\ \ \ \ {\bf attributes} type A1, ..., AM\\
\ \ \ \ \ \ \vdots\\
\ \ \ {\bf rule} $[[$Y$]$ {\bf extends} $|$ $[$Y$]$ {\bf overrides}$]$ Z \{\\
\ \ \ \ \ \ X ::= X$_{11}$ X$_{12}$ ... X$_{1p}$ {\bf compute} \{\\
\ \ \ \ \ \ \ \ \ \ semantic functions \}\\
\ \ \ \ \ \ \ \ \vdots\\
\ \ \ \ \ \ $|$\\
\ \ \ \ \ \ \ \ X$_{r1}$ X$_{r2}$ ... X$_{rt}$ {\bf compute} \{\\
\ \ \ \ \ \ \ \ \ \ semantic functions \} \\
\ \ \ \ \ \ ;\\
\ \ \ \ \ \ \}\\
\ \ \ \ \ \ \vdots\\
\ \ \ \ {\bf method} $[[$N$]$ {\bf overrides} $|$ $[$N$]$ {\bf extends}$]$ M \{\\
\ \ \ \ \ \ operations on semantic domains\\
\ \ \ \ \ \ \}\\
\ \ \ \ \vdots\\
\ \ \} \\
\hline
\end{tabular}
\end{center}
\normalsize
\vspace{-5mm}
\end{table}

Lexical specifications for EasyTime in LISA (Fig.~\ref{pic:LISA_GUI}) are similar to those used in other compiler-generators,
and are obtained from EasyTime concrete syntax (Table~\ref{tab:tab17}). Note that in the rule part of LISA specifications the terminal symbols that are defined by regular expressions in the lexical part are denoted with symbol \# (e.g., \#Id, \#Int).
EasyTime concrete syntax is derived from EasyTime abstract syntax (Table~\ref{tab:X}). The process of transforming abstract syntax into concrete syntax is straightforward, and presented in~\cite{Fister:2011}.
Semantic rules are written in LISA as regular Java assignment statements and are attached to a particular syntax rule. Hence, the rule part in LISA (Table~\ref{tab:tab30}) specifies the BNF production as well as the attribute computations attached to this production.
Since the theory about attribute grammars is a standard topic of compiler science, it is assumed that a reader has a basic knowledge about attribute grammars ~\cite{Knuth:1968,Paakki:1995}.

\begin{table}[htb]
\caption{The concrete syntax of EasyTime}
\label{tab:tab17}
\vspace{-5mm}
\footnotesize
\begin{center}
\begin{tabular}{ | l  l  l | }
\hline
  PROGRAM & ::= & AGENTS DECS MES\_PLACES \\
  AGENTS & ::= & AGENTS AGENT \textbar\ $\varepsilon$ \\
  AGENT & ::= & \#Int auto \#ip ; \textbar\  \#Int manual \#file ; \\
  DECS & ::= & DECS DEC \textbar\ $\varepsilon$ \\
  DEC & ::= & var \#Id $:=$ \#Int ; \\
  MES\_PLACES & ::= & MES\_PLACE MES\_PLACES \textbar\ MES\_PLACE \\
  MES\_PLACE & ::= & mp[ \#Int ] $->$ agnt [ \#Int ] \{ STMTS \} \\
  STMTS & ::= & STMT STMTS \textbar\ STMT  \\
  STMT & ::= & dec \#Id ; \textbar\ upd \#Id ; \textbar\ \#Id $:=$ EXPR ; \textbar\ ( LEXPR ) $->$ STMT \\
  LEXPR & ::= & true \textbar\ false \textbar\ EXPR == EXPR \textbar\ EXPR != EXPR \\
  EXPR & ::= & \#Int \textbar\ \#Id \\
\hline
\end{tabular}
\end{center}
\normalsize
\vspace{-5mm}
\end{table}

\subsection{Translation scheme from denotational semantics to attribute grammars}

The most difficult part of transforming formal EasyTime specifications into LISA specifications, consists of mapping denotational semantics into attribute grammars. This mapping can be described in a systematic manner, and can also be used for the implementation of other DSLs (e.g.,~\cite{Lukovic:2011}). It consist of the following steps similar to the translation scheme from natural semantics into attribute grammars \cite{Attali1994}:

\begin{enumerate}
  \item Identification of syntactic and semantic domains in each semantic function of denotational semantics. Identified syntactic domains must have their counterparts in non-terminals of concrete syntax. Identified semantic domains must be represented appropriately, with suitable data structures (ty\-pes) in chosen programming language.
  \item Identification of inherited and synthesized attributes for each non-terminal derived in step 1. Semantic argument, which is an input parameter in semantic function, is represented as inherited attribute, while an output parameter is represented as synthesized attribute. According to \cite{Knuth:1968}, the starting non-terminal should not have inherited attributes. Whilst LISA automatically infers whether an attribute is inherited or synthesized \cite{Knuth:1968}, the type of attribute must be specified (Fig.~\ref{pic:LISA_GUI}).
  \item For all identified attributes attached to a particular non-terminal's, semantic equations need to be developed that are in conformance to semantic equations from denotational semantics. In particular, semantic equations need to be written for each synthesized attribute of the left-hand side non-terminal and for each inherited attribute attached to non-terminals of the right-hand side. This rule is applied to every production of a concrete syntax. In this step the whole semantic equation is not yet written, only the existence of such an equation is identified.
  \item In the productions of concrete syntax certain new non-terminals appear, which are consequences of transformation of abstract syntax into concrete syntax. These non-terminals also carry information that are needed for computations. In this step such non-terminals are identified and attached attributes are classified into inherited and synthesized.
  \item Finalizing semantics for all identified semantic equations. These semantic equations need to be in conformance to denotational semantics, and require careful examination of semantic functions of denotational semantics (e.g., $\mathcal{CP}$, $\mathcal{A}$, $\mathcal{D}$, $\mathcal{CM}$, $\mathcal{CS}$,
$\mathcal{CB}$, and $\mathcal{CA}$ from Table~\ref{tab:tab6}). This step is most demanding.
  \item In code generation, certain additional tests are usually performed, which are sometimes non-described in formal semantics, in order to be on a proper abstraction level. For example, only declared variables can be used in expressions and commands of a language under development. Such additional tests require that new attributes are defined to carry the results of tests, as well as existing attributes being propagated to appropriate constructs (e.g., expressions, commands). An attribute grammar is finalized during this step.
\end{enumerate}

Note that the presented guidelines are general and not restricted to a particular class of attribute grammars~\cite{Knuth:1968,Paakki:1995} (e.g., S-attributed, L-attributed, ordered attribute grammar, absolutely non-circular attribute grammar). Actually, the class of obtained attribute grammar can be identified only after the translation has been completely performed.

\subsection{Translation scheme from EasyTime formal semantics to LISA}
When applying the aforementioned rules to EasyTime, the following results are obtained after each step.

Step 1:

The following non-terminals from Table \ref{tab:tab17} represent syntactic domains (Table~\ref{tab:X}):
PROGRAM $\in$ \textbf{Pgm}, MES\_PLACES $\in$ \textbf{MeasPlace}, DECS $\in$ \textbf{Dec}, AGENTS $\in$ \textbf{Adec},
STMTS $\in$ \textbf{Stm}, etc. Semantic domains (Table~\ref{tab:Y}) such as \textbf{Integer}, \textbf{Truth-Value}, \textbf{Code} have direct counterparts with Java types: int, boolean, and String. While semantic domains which are functions (e.g., \textbf{State}, \textbf{Agents}) can be modeled with Java  Hashtable type.
For example, from Figure~\ref{pic:LISA_GUI} we can notice that attribute $inState$, which represents function \textbf{State}, is of type Hashtable. Using methods such as
$put()$, $get()$, and $containsKey()$ we can respectively insert a new variable, obtain a variable's value, and check if the variable is declared.
Other semantic domains (e.g., cartesian product) can be modeled easily with a Java rich type system. Hence, in LISA the type of attributes regarding an attribute grammar can be any valid pre-defined or user-defined Java type. An example of auxiliary operations on semantic domains (e.g., Hashtable), is presented in \cite{Fister:2011a}.

\newpage
Step 2:

From $\mathcal{CP}:\textbf{Pgm} \rightarrow \textbf{Runners} \rightarrow$ \textbf{Code} $\times$ \textbf{Integer} $\times$ \textbf{DataBase}  (Table \ref{tab:tab6}) it can be concluded that to non-terminal PROGRAM one inherited (representing a parameter of type \textbf{Runners}) and three synthesized attributes (representing parameters of \textbf{Code}, \textbf{Integer}, and \textbf{DataBase}) need to be attached. However, the starting non-terminal should not have inherited attributes~\cite{Knuth:1968,Paakki:1995}. From the definition of semantic function $\mathcal{CP}$ (Table~\ref{tab:tab6}) it can be noticed that the input parameter of type \textbf{Runners} are only needed to create a database. Hence, both parameters (of type \textbf{Runners} and \textbf{DataBase}) can be omitted from LISA specifications, and its functionality can be externally implemented. Moreover, it was decided to represent both the generated code and the identification number of the virtual machine, where the code is going to be executed, as a string \textbf{"}(Code, Integer)\textbf{"}. Hence, only one synthesized attribute, PROGRAM.code, is attached to starting non-terminal PROGRAM.\\
From $\mathcal{A}:\textbf{Adec}\rightarrow \textbf{Agents} \rightarrow$ \textbf{Agents} (Table \ref{tab:tab6}) it can be concluded that one inherited and one synthesized attribute need to be attached to non-terminal AGENTS. For this purpose AGENTS.inAG is an inherited attribute, and AGEN\-TS.outAG a synthesized attribute. Both attributes are of type Hashtable since semantic domain \textbf{Agents} is a function, which can be modeled as a Hashtable.\\
From $\mathcal{D}:\textbf{Dec}\rightarrow \textbf{State} \rightarrow$ \textbf{State} (Table \ref{tab:tab6}) it can be concluded that one inherited and one synthesized attributes need to be attached to non-terminal DECS. For this purpose DECS.inState is inherited attribute, and DECS.outState a synthesized attribute. Both attributes are of type Hashtable since semantic domain \textbf{State} is a function, which can be modeled as a Hashtable.\\
From $\mathcal{CM}:\textbf{MeasPlace} \rightarrow \textbf{Agents} \rightarrow$ \textbf{Code} $\times$ \textbf{Integer} (Table \ref{tab:tab6}) it can be concluded that one inherited and two synthesized attributes need to be attached to non-terminal MES\_PLACES. Again, it was decided to represent both, a generated code and the identification number of virtual machine, as a string. For this purpose MES\_\-PLACES.inAG is an inherited attribute and MES\_PLACES.code is a synthesized attribute.\\
From $\mathcal{CS}:\textbf{Stm} \rightarrow$ \textbf{Agents} $\times$ \textbf{Integer} $\rightarrow$ \textbf{Code} (Table \ref{tab:tab6}) it can be concluded that two inherited and one synthesized attribute need to be attached to non-terminal STMTS. For this purpose STMTS.inAG and STMTS.n are inherited attributes of type Hash\-table and int, respectively. The attribute STMTS.code is a synthesized attribute of type String. The attributes, inherited and synthesized, attached to the appropriate non-terminals are collated in Table~\ref{tab:tab18}.

\begin{table}[htb]           
\caption{Attributes of non-terminals representing syntactic domains from EasyTime formal semantics}
\label{tab:tab18}
\vspace{-5mm}
\footnotesize
\begin{center}
\begin{tabular}{ | l | l | l | }
\hline
X          &   Inherited(X)      &   Synthesized(X)\\
\hline
PROGRAM    &          &       code\\
AGENTS     &   inAG              &      outAG\\
DECS       &   inState           &      outState\\
MES\_PLACES &   inAG              &      code\\
STMTS      &   inAG, n           &      code\\
\hline
\end{tabular}
\end{center}
\normalsize
\vspace{-5mm}
\end{table}

Step 3:

In this step semantic equations are given for each synthesized attribute of the left-hand side non-terminal, and for each inherited attribute for the right-hand side non-terminal. This procedure is applied to each production in the context-free grammar (Table \ref{tab:tab17}). The LISA specification fragment as illustrated in Table~\ref{tab:tab19} indicates, which semantic equations need to be developed. Let us explain the process for the first production. Since the non-terminal PROGRAM, left-hand side non-terminal, has only one synthesized attribute $code$ (Table \ref{tab:tab18}) only one semantic equation must be defined (PROGRAM.code = ...;). Other non-terminals (AGENTS, DECS, MES\_PLACES) in the first production are on the right hand side and hence only inherited attributes attached to those non-terminals must be defined (AGENTS.inAG = ...; DECS.\-inState = ...; MES\_\-PLACES.\-inAG = ...;). Note that the order of these semantic equations is irrelevant~\cite{Knuth:1968,Paakki:1995}.

\begin{table}[htb]           
\caption{Semantic equations under development that are obtained after Step 3}
\label{tab:tab19}
\vspace{-5mm}
\footnotesize
\begin{center}
\begin{tabular}{ | l  l  l | }
\hline
PROGRAM & ::= & AGENTS DECS MES\_PLACES  compute \{ \\
 & &            \ \ \ \ \ \ \ \ AGENTS.inAG = ...; \\
 & &            \ \ \ \ \ \ \ \ DECS.inState = ...; \\
 & &            \ \ \ \ \ \ \ \ MES\_PLACES.inAG = ...; \\
 & &            \ \ \ \ \ \ \ \ PROGRAM.code = ...; \}; \\
 & & \\
AGENTS & ::= & AGENTS  AGENT compute \{ \\
 & &       \ \ \ \ \ \ \ \ AGENTS[1].inAG = ...; \\
 & &       \ \ \ \ \ \ \ \ AGENTS[0].outAG = ...; \};\\
 & & \\
DECS & ::= & DECS DEC compute \{ \\
 & &        \ \ \ \ \ \ \ \ DECS[1].inState = ...; \\
 & &        \ \ \ \ \ \ \ \ DECS[0].outState = ...; \}; \\
 & & \\
MES\_PLACES & ::= & MES\_PLACE MES\_PLACES compute \{ \\
 & &            \ \ \ \ \ \ \ \ MES\_PLACES[1].inAG = ...; \\
 & &            \ \ \ \ \ \ \ \ MES\_PLACES[0].code = ...; \}; \\
 & & \\
STMTS & ::= & STMT STMTS compute \{ \\
 & &            \ \ \ \ \ \ \ \ STMTS[1].n = ...; \\
 & &            \ \ \ \ \ \ \ \ STMTS[1].inAG = ...; \\
 & &            \ \ \ \ \ \ \ \ STMTS[0].code = ...; \}; \\
\hline
\end{tabular}
\end{center}
\normalsize
\vspace{-5mm}
\end{table}

\newpage
Step 4:

From step 3, it can be identified the following non-terminals, which appears in concrete syntax (Table \ref{tab:tab17}) and were unidentified in steps 1 - 3: AGENT, DEC, MES\_\-PLACE, and STMT (Table~\ref{tab:tab24}). If the structure of these non-terminals is simple (e.g., AGENT, DEC) then attributes attached to these non-terminals carried only synthesized attributes representing mostly lexical values (Table~\ref{tab:tab25}). Semantic equations can be derived immediately for those attributes. On the other hand, some non-terminals might be complex (e.g., MES\_\-PLACE, STMT) and inherited attributes attached to these non-terminals are also needed. The attributes might be similar to those attributes attached to other non-terminals in productions, where new non-terminals appear (Table~\ref{tab:tab18}). Moreover, semantic equations may no longer be simple (Table~\ref{tab:tab25}). For example, attributes attached to non-terminals MES\_\-PLACE and STMT (Table~\ref{tab:tab24}) are the same as those attached to non-terminals STMTS and MES\_PLACES, respectively (Table~\ref{tab:tab18}). However, due to the semantics of the update statement (Table~\ref{tab:tab6}) another attribute $y$ is attached to the non-terminal STMT (Table~\ref{tab:tab24}).

\begin{table}[htb]           
\caption{Attributes for additional non-terminals}
\label{tab:tab24}
\vspace{-5mm}
\footnotesize
\begin{center}
\begin{tabular}{ | l | l | l | }
\hline
X          &   Inherited(X)  &       Synthesized(X) \\
\hline
AGENT      &                 &      number, type, file\_ip \\
DEC        &                 &      name, value \\
MES\_PLACE  &  inAG           &         code \\
STMT       &  inAG, n        &         code, y \\
\hline
\end{tabular}
\end{center}
\normalsize
\vspace{-5mm}
\end{table}

\begin{table}[htb]           
\caption{Semantic equations for additional non-terminals}
\label{tab:tab25}
\vspace{-5mm}
\footnotesize
\begin{center}
\begin{tabular}{ | l  l  l | }
\hline
AGENT & ::= & \#Int auto \#ip \; compute \{ \\
 & &  \ \ \ \ \ \ \ \ AGENT.number = Integer.valueOf(\#Int[0].value()).intValue(); \\
 & &  \ \ \ \ \ \ \ \ AGENT.type = "auto"; \\
 & &  \ \ \ \ \ \ \ \ AGENT.file\_ip = \#ip.value(); \}; \\
 & & \\
DEC & ::= & var \#Id \:\=  \#Int \; compute \{ \\
 & &  \ \ \ \ \ \ \ \ DEC.name = \#Id.value(); \\
 & &  \ \ \ \ \ \ \ \ DEC.value = Integer.valueOf(\#Int.value()).intValue(); \}; \\
 & & \\
MES\_PLACES & ::= & MES\_PLACE MES\_PLACES compute \{ \\
 & &  \ \ \ \ \ \ \ \ MES\_PLACE.inAG = ...; \}; \\
 & & \\
MES\_PLACE & ::= & mp [ \#Int ] $->$  agnt [ \#Int ] \{ STMTS \} compute \{ \\
 & &  \ \ \ \ \ \ \ \ MES\_PLACE.code= ...; \}; \\
 & & \\
STMTS & ::= & STMT STMTS compute \{ \\
 & &  \ \ \ \ \ \ \ \ STMT.n = ...; \\
 & &  \ \ \ \ \ \ \ \ STMT.inAG = ...; \}; \\
 & & \\
STMT & ::= & upd  \#Id \; compute \{ \\
 & &  \ \ \ \ \ \ \ \ STMT.y =  ...; \\
 & &  \ \ \ \ \ \ \ \ STMT.code = ...; \}; \\
\hline
\end{tabular}
\end{center}
\normalsize
\vspace{-5mm}
\end{table}

Step 5:

The reasoning of this step is only explained for semantic functions $\mathcal{A}$ and $\mathcal{CM}$ (Table \ref{tab:tab6}), which are translated into attributes for non-terminals AGENTS, AGENT, MES\_PLACES, and MES\_PLACE (Tables ~\ref{tab:tab18} and ~\ref{tab:tab24}). For other semantic functions the reasoning is similar.
The semantic equation $\mathcal{A} \lsem A_{1};A_{2}\rsem ag$ = $\mathcal{A} \lsem A_{2} \rsem$ $(\mathcal{A} \lsem A_{1} \rsem ag)$ (Table \ref{tab:tab6}) constructs $ag \in Agents$, which is a function from an integer, denoting an agent, into an agent's type (manual or auto), and an agent's ip or agent's file. This function is described in LISA  as presented in Table~\ref{tab:tab25a}. From Table~\ref{tab:tab25a} it can be noticed how the attribute $outAG$, which represents the $ag \in Agents$, is constructed simply by the calling method $insert()$. The method $insert()$ will insert a new agent with a particular number, type, and file\_ip into the Hashtable. Note also, how the missing equations from Step 3 have been developed.
The net effect is that we are constructing a list, more precisely a hash table, of agents where we are recording the agent's number ($AGENT.number$), the agents's type ($AGENT.type$), and the agent's ip or file ($AGENT.file\_ip$) (see Step 4). The complete LISA specifications for semantic function $\mathcal{A}$, is shown in Algorithm \ref{alg:agent_lisa}.

\begin{table}[htb]           
\caption{Semantic equation for AGENTS}
\label{tab:tab25a}
\vspace{-5mm}
\footnotesize
\begin{center}
\begin{tabular}{ | l  l  l | }
\hline
AGENTS & ::= & AGENTS  AGENT compute \{ \\
 & & \ \ \ \ \ \ \ \ AGENTS[1].inAG = AGENTS[0].inAG; \\
 & & \ \ \ \ \ \ \ \ AGENTS[0].outAG = insert(AGENTS[1].outAG, \\
 & & \ \ \ \ \ \ \ \ new Agent(AGENT.number, AGENT.type, AGENT.file\_ip)); \\
 & & \ \ \ \ \ \ \ \ \} \\
 & & \ \ \ \ $\mid$ epsilon compute \{ \\
 & & \ \ \ \ \ \ \ \ AGENTS.outAG = AGENTS.inAG; \\
 & & \ \ \ \ \ \ \ \ \}; \\
\hline
\end{tabular}
\end{center}
\normalsize
\vspace{-5mm}
\end{table}

\begin{algorithm}[tbh]
\caption{Translation of Agents into LISA specifications}
\label{alg:agent_lisa}
\scriptsize
\begin{algorithmic}[1]
\STATE rule Agents \{
\STATE \ \ \ \ AGENTS ::= AGENTS  AGENT compute \{
\STATE \ \ \ \ \ \ \ \ AGENTS[1].inAG = AGENTS[0].inAG;
\STATE \ \ \ \ \ \ \ \ AGENTS[0].outAG = insert(AGENTS[1].outAG,
\STATE \ \ \ \ \ \ \ \ \ new Agent(AGENT.number, AGENT.type, AGENT.file\_ip));
\STATE \ \ \ \ \}
\STATE \ \ \ \ $|$ epsilon compute \{
\STATE \ \ \ \ \ \ \ \ AGENTS.outAG = AGENTS.inAG;
\STATE \ \ \ \ \};
\STATE \}
\STATE rule AGENT \{
\STATE \ \ \ \ AGENT ::= \#Int manual \#file \; compute \{
\STATE \ \ \ \ \ \ \ \ AGENT.number = Integer.valueOf(\#Int[0].value()).intValue();
\STATE \ \ \ \ \ \ \ \ AGENT.type = "manual";
\STATE \ \ \ \ \ \ \ \ AGENT.file\_ip = \#file.value();
\STATE \ \ \ \ \};
\STATE \ \ \ \ AGENT ::= \#Int auto \#ip \; compute \{
\STATE \ \ \ \ \ \ \ \ AGENT.number = Integer.valueOf(\#Int[0].value()).intValue();
\STATE \ \ \ \ \ \ \ \ AGENT.type = "auto";
\STATE \ \ \ \ \ \ \ \ AGENT.file\_ip = \#ip.value();
\STATE \ \ \ \ \};
\STATE \}
\end{algorithmic}
\normalsize
\end{algorithm}

The reasoning for the semantic function $\mathcal{CM}$ is done in a similar manner. The semantic equation $\mathcal{CM} \lsem M_{1}; M_{2} \rsem ag$ = $\mathcal{CM} \lsem M_{1} \rsem ag: \mathcal{CM} \lsem M_{2} \rsem ag$ (Table \ref{tab:tab6}) translates the first construct $M_1$ into code before performing the translation of the second construct $M_2$. This function is described in LISA, as represented in Table~\ref{tab:tab25b}, with the following meaning: The code for the first construct $\mathit{MES\_PLACE}$ is simply concatenated with the code from the second construct $MES\_PLACES[1]$.

\begin{table}[htb]           
\caption{Semantic equation for MES\_PLACES}
\label{tab:tab25b}
\vspace{-5mm}
\footnotesize
\begin{center}
\begin{tabular}{ | l  l  l | }
\hline
MES\_PLACES & ::= & MES\_PLACE MES\_PLACES compute \{ \\
 & & \ \ \ \ \ \ \ \ MES\_PLACES[0].code = MES\_PLACE.code + \\
 & & \ \ \ \ \ \ \ \ "$\backslash$ n"  + MES\_PLACES[1].code; \}; \\
MES\_PLACES & ::= & MES\_PLACE compute \{ \\
 & & \ \ \ \ \ \ \ \ MES\_PLACES.code = MES\_PLACE.code \}; \\
\hline
\end{tabular}
\end{center}
\normalsize
\vspace{-5mm}
\end{table}

The semantic equation
$\mathcal{CM}\lsem \textbf{mp}[n_{1}]\rightarrow\textbf{agnt}[n_{2}] S \rsem ag$ = $(WAIT\ i:\mathcal{CS} \lsem S \rsem$ $(ag,n_{2}), n_{1})$ (Table \ref{tab:tab6}) is described in LISA, as presented in Table~\ref{tab:tab26}.

\begin{table}[htb]           
\caption{Semantic equation for MES\_PLACE}
\label{tab:tab26}
\vspace{-5mm}
\footnotesize
\begin{center}
\begin{tabular}{ | l  l  l | }
\hline
MES\_PLACE & ::= & mp [ \#Int ] $->$  agnt [ \#Int ] \{ STMTS \} compute \{ \\
 & & \ \ \ \ \ \ \ \ MES\_PLACE.code= "(WAIT i " + STMTS.code + \\
 & & \ \ \ \ \ \ \ \ ", " + \#Int[0].value() + ")"; \}; \\
\hline
\end{tabular}
\end{center}
\normalsize
\vspace{-5mm}
\end{table}

However, in this step the undefined semantic equations from steps 3 and 4 also need to be developed (Table~\ref{tab:tab27}). For example, a list of agents ($inAG$) needs to be propagated.

\begin{table}[htb]           
\caption{Developing undefined semantic equations for MES\_PLACES}
\label{tab:tab27}
\vspace{-5mm}
\footnotesize
\begin{center}
\begin{tabular}{ | l  l  l | }
\hline
MES\_PLACES & ::= & MES\_PLACE MES\_PLACES compute \{ \\
 & & \ \ \ \ \ \ \ \ MES\_PLACE.inAG = MES\_PLACES[0].inAG; \\
 & & \ \ \ \ \ \ \ \ MES\_PLACES[1].inAG = MES\_PLACES[0].inAG; \\
 & & \ \ \ \ \ \ \ \ ...  \}; \\
MES\_PLACES & ::= & MES\_PLACE compute \{ \\
 & & \ \ \ \ \ \ \ \ MES\_PLACE.inAG = MES\_PLACES.inAG; \\
 & & \ \ \ \ \ \ \ \ ... \}; \\
\hline
\end{tabular}
\end{center}
\normalsize
\vspace{-5mm}
\end{table}

Step 6:

Easytime also uses variables in statements, and additional checks must be performed if only declared variables appear in expressions and statements. For this reason an additional attribute $ok$ of type boolean has been introduced into the specifications. Moreover, to be able to check if a variable is declared, it is necessary to propagate attribute $inState$ into the measuring places, statements, and expressions. The complete LISA specifications for MES\_PLACE are shown in Algorithm \ref{alg:mp_lisa} also using attributes $ok$ and $inState$.

\begin{table}[htb]           
\caption{Semantic equations for the starting production}
\label{tab:tab28}
\footnotesize
\vspace{-5mm}
\begin{center}
\begin{tabular}{ | l  l  l | }
\hline
 PROGRAM & ::= & AGENTS DECS MES\_PLACES  compute \{ \\
 & &  \ \ \ \ \ \ \ \ AGENTS.inAG = new Hashtable(); \\
 & &  \ \ \ \ \ \ \ \ DECS.inState = new Hashtable(); \\
 & &  \ \ \ \ \ \ \ \ MES\_PLACES.inAG = AGENTS.outAG; \\
 & &  \ \ \ \ \ \ \ \ MES\_PLACES.inState = DECS.outState; \\
 & &  \ \ \ \ \ \ \ \ PROGRAM.code = MES\_PLACES.ok ? "$\backslash$ n" + \\
 & &  \ \ \ \ \ \ \ \ MES\_PLACES.code + "$\backslash$ n" : "ERROR"; \}; \\
\hline
\end{tabular}
\end{center}
\vspace{-5mm}
\normalsize
\end{table}

\begin{algorithm}[tbh]
\caption{Translation of MES\_PLACE into LISA specifications}
\label{alg:mp_lisa}
\scriptsize
\begin{algorithmic}[1]
\STATE rule Mes\_places \{
\STATE \ \ \ \ MES\_PLACES ::= MES\_PLACE MES\_PLACES compute \{
\STATE \ \ \ \ \ \ \ \ MES\_PLACE.inAG = MES\_PLACES[0].inAG;
\STATE \ \ \ \ \ \ \ \ MES\_PLACES[1].inAG = MES\_PLACES[0].inAG;
\STATE \ \ \ \ \ \ \ \ MES\_PLACE.inState = MES\_PLACES[0].inState;
\STATE \ \ \ \ \ \ \ \ MES\_PLACES[1].inState = MES\_PLACES[0].inState;
\STATE \ \ \ \ \ \ \ \ MES\_PLACES[0].ok = MES\_PLACE.ok \&\& MES\_PLACES[1].ok;
\STATE \ \ \ \ \ \ \ \ MES\_PLACES[0].code = MES\_PLACE.code + "$\backslash$n" + MES\_PLACES[1].code;
\STATE \ \ \ \ \};
\STATE MES\_PLACES ::=  MES\_PLACE compute \{
\STATE \ \ \ \ \ \ \ \ MES\_PLACE.inAG = MES\_PLACES.inAG;
\STATE \ \ \ \ \ \ \ \ MES\_PLACE.inState = MES\_PLACES.inState;
\STATE \ \ \ \ \ \ \ \ MES\_PLACES.ok = MES\_PLACE.ok;
\STATE \ \ \ \ \ \ \ \ MES\_PLACES.code = MES\_PLACE.code;
\STATE \ \ \ \ \};
\STATE \}
\STATE rule MES\_PLACE \{
\STATE \ \ \ \ MES\_PLACE ::= mp $\backslash[$ \#$\mathit{Int}$ $\backslash]$ $\backslash-\backslash>$ $\mathit{agnt}$
$\backslash[$ \#$\mathit{Int}$ $\backslash]$ $\backslash\{$ STMTS $\backslash\}$ compute \{
\STATE \ \ \ \ \ \ \ \ STMTS.inAG = MES\_PLACE.inAG;
\STATE \ \ \ \ \ \ \ \ STMTS.inState = MES\_PLACE.inState;
\STATE \ \ \ \ \ \ \ \ STMTS.n = Integer.valueOf(\#Int[1].value()).intValue();
\STATE \ \ \ \ \ \ \ \ MES\_PLACE.ok = STMTS.ok;
\STATE \ \ \ \ \ \ \ \ MES\_PLACE.code = "(WAIT i " + STMTS.code + ", " + \#Int[0].value() + ")";
\STATE \ \ \ \ \};
\STATE \}
\end{algorithmic}
\normalsize
\end{algorithm}

Semantic equations for other production are obtained in a similar manner. Let us conclude this example by finalizing semantic equations for the starting production (see also Table~\ref{tab:tab19}). The initial hash table for agents ($AGENTS.inAG$) and declarations ($DECS.inState$) are empty (Table~\ref{tab:tab28}). Agents and declarations are constructed after visiting the subtrees represented by the non-terminals $AGENTS$ and $DECS$, and stored into attributes $AGENTS.outAG$ and $DECS.$ $outState$, that are passed to the subtree represented by the non-terminal $MES\_$ $PLACES$. If all the syntactic constraints are satisfied ($MES\_PLACES.ok==true$), then the generated code is equal to a code produced by the subtree represented by the non-terminal $MES\_PLACES$.

\section{Operation}

Local organizers of sporting competitions were faced with two possibilities before developing EasyTime:
\begin{itemize}
  \item to rent a specialized company to measure time,
  \item to measure time manually.
\end{itemize}
The former possibility is expensive, whilst the latter can be very unreliable. However, both objectives (i.e. inexpensiveness and reliability), can be fulfilled by EasyTime. On the other hand, producers of measuring devices usually deliver their units with software for the collecting of events into a database. Then these events need to be post-processed (batch processed) to get the final results of the competitors. Although this batch-processing can be executed whenever the organizer desires, each real-time application requests online processing. Fortunately, EasyTime enables both kinds of event processing.

In order to use the source program written in EasyTime by the measuring system, it needs to be compiled. Note that the code
generation \cite{Aho:1972} of a program in EasyTime is performed only if the parsing is finished successfully. Otherwise the compiler
prints out an error message and stops. For each of measuring places individually, the code is automatically generated by strictly following the rules, as defined in Section 3. An example of the generated code from the Algorithm~\ref{alg:prog} for the controlling of measurements, as illustrated by Fig.~\ref{pic:slika_1}, is presented in Table~\ref{tab:tab10}. Note that the generated code is saved
into a database. The meaning of the particular instructions of virtual machine (e.g., WAIT, FETCH, STORE), is explained in Table~\ref{tab:am}.

\begin{table}[htb]
\caption{Translated code for the EasyTime program in Algorithm~\ref{alg:prog}}
\label{tab:tab10}
\begin{center}
\vspace{-5mm}
\small
\begin{tabular}{ | l | }
\hline
(WAIT i FETCH accessfile("abc.res") STORE SWIM \\
FETCH ROUND1 DEC STORE ROUND1, 1) \\ \\
(WAIT i FETCH accessfile("abc.res") STORE TRANS1, 2) \\ \\
(WAIT i FETCH connect(192.168.225.100) STORE INTER2 \\
FETCH ROUND2 DEC STORE ROUND2 \\
PUSH 0  FETCH ROUND2  EQ BRANCH( FETCH \\
connect(192.168.225.100) STORE BIKE, NOOP), 3) \\ \\
(WAIT i FETCH connect(192.168.225.100) STORE INTER3 \\
PUSH 55  FETCH ROUND3  EQ BRANCH( FETCH \\
connect(192.168.225.100) STORE TRANS2, NOOP) \\
FETCH ROUND3 DEC STORE ROUND3 \\
PUSH 0  FETCH ROUND3  EQ BRANCH( FETCH \\
connect(192.168.225.100) STORE RUN, NOOP), 4) \\
\hline
\end{tabular}
\vspace{-5mm}
\normalsize
\end{center}
\end{table}

\begin{figure}[htb]
\vspace{-5mm}
    \begin{center}
        \includegraphics [scale=0.85]{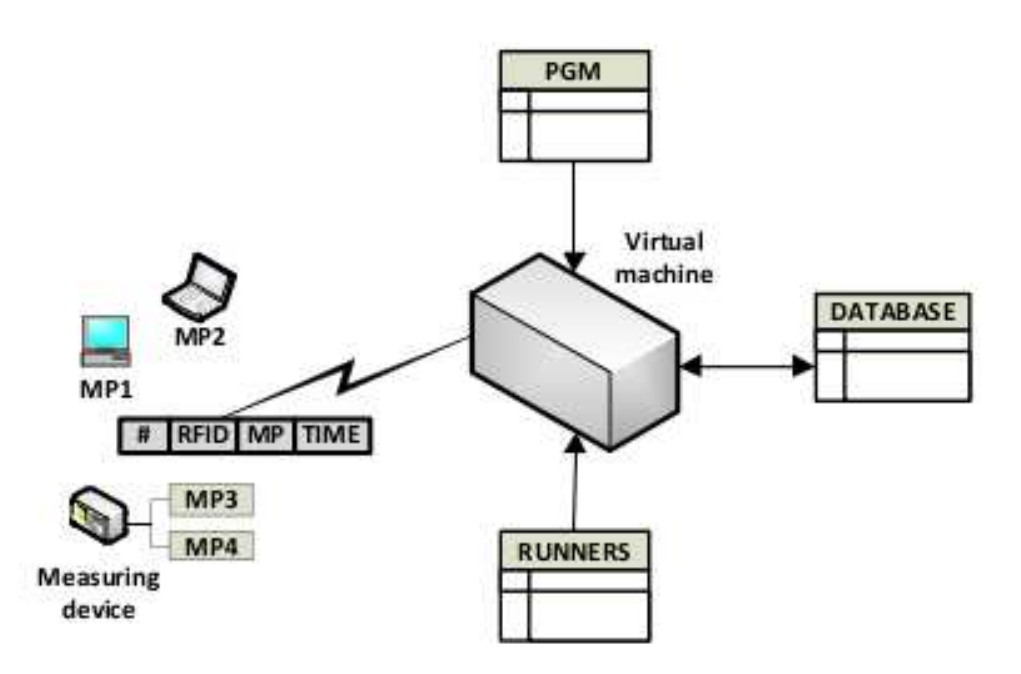}      %
        \caption{Executable environment of a program in EasyTime}
        \label{pic:slika_3}
    \end{center}
\vspace{-5mm}
\end{figure}

As a matter of fact, the generated code is dedicated to the control of an agent by writing the events received from the measuring
devices, into the data\-base. Normally, the program code is loaded from the database only once. That is, only an
interpretation of the code could have any impact on the performance of a measuring system. Because this interpretation is not time consuming, it cannot degrade the performance of the system. On the other hand, the precision of measuring time is handled by the measuring device and is not changed by the processing of events. In fact, the events can be processed as follows:
\begin{itemize}
  \item batch: manual mode of processing, and
  \item online: automatic mode of processing.
\end{itemize}
The agent reads and writes the events that are collected in a text file, when the first mode of processing is assumed. Typically, events captured by a computer timer are processed in this mode. Here, the agent looks for an existence of the event text file that is
configured in the agent statement. If it exists, the batch processing is started. When the processing is finished, the text file is
archived and then deleted. The online processing is event oriented, i.e. each event generated by the measuring device is
processed in time. In both modes of processing, the agent works with the program PGM, the runner table RUNNERS, and the results table DATABASE, as can be
seen in Fig.~\ref{pic:slika_3}. An initialization of the virtual machine is performed when the agent starts. The initialization consists of loading the program code from PGM. That is, the code is loaded only once. At the same time, the variables are initialized on starting values.

In order to ensure the reliability of Easytime in practice, competitors are not allowed to go directly from swimming to running, because the course is complex and the competitor must to go through both transition areas. In the case that a competitor skips over the next discipline, the referees disqualify him/her immediately. Actually, EasyTime is only of assistance to referees. All misuses of the triathlons rules do not have any impact on its operation.

\begin{figure*}[htb]      
    \begin{center}
        \includegraphics [scale=0.6] {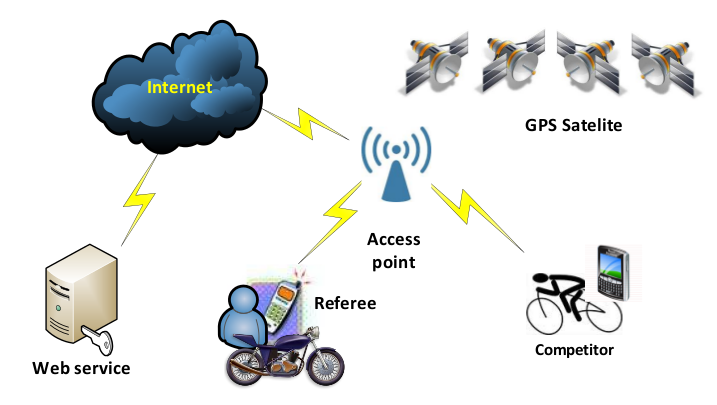}  
        \caption{Proposed system for drafting detection in triathlons}
        \label{pic:sys}
    \end{center}
\vspace{-5mm}
\end{figure*}

After the development of EasyTime another demand has arisen - drafting detection in triathlons.
This problem is especially expressive in cycling, where competitors wishing to improve their results ride their cycles within close-knit groups. In this way, competitors achieve a higher speed and save energy for later efforts. Typically, within such groups of competitors the hardest work is performed by the leading competitor because he needs to overcome on air resistance. At the same time, other competitors may take a rest. Actually, the drafting violation arises when one competitor rides behind the other closer than 7 meters for more than 20 seconds. Interestingly, this phenomenon is only pursued during long-distance triathlons, whilst drafting is allowed over short-distances.
Any competitor who violates this drafting rule is punished by the referees with 5 minutes of elimination from the cycling race.  The referees observe the race from motorcycles and determine the drafting violations according to their feelings. In this sense only, this assessment is very subjective. On the other hand, the referees can control one competitor a time. Consequently, an automatic system is needed for detecting drafting violations during triathlons.
A drafting detection system is proposed in order to track this violation. This system is based on smart-phones because these incorporate the following features: information access via wireless networks and GPS navigation. Smart-phones need to be borne by competitors on their bicycles (Fig.~\ref{pic:sys}). These determine information about competitor current GPS positions and transmit these over wireless modems to a web-service. From the positions of all competitors the web-service calculates whether a particular competitor is violating the drafting rule. In addition, these violations can be tackled by the referees on motorcycles using smart-phones.

Normally, the organizers of triathlons demand the integration of EasyTime within the system for drafting violation. At a glance, this integration can be performed at the computer-system level, i.e., the mobile agent is added to the existing EasyTime agents. This mobile agent acts as a web-service and runs on an application server. Like EasyTime, it uses its own database. Each record in this database represents a competitor's current GPS position that can be defined as tuple $\langle \#,x,y,z,t,l \rangle$, where $\#$ denotes the competitor's starting number, $x,y,z$ his current position within the coordinate system UTM, $t$ the registration time in the mobile device, and $l$ the calculated path-length. This length $l$ is obtained by projecting the current position of the $\#$-th competitor on the line that connects the points gained by tracking the cycling course with a precise GPS device, at each second. This has an impact on the competitor's current position, from which the distance is calculated to the competitor in front of him.
At the moment, both systems run on the same server separately.  However, further development of a wireless technology and pervasive computing~\cite{Weiser:1991} indicates that EasyTime should have the ability to run on an application server as well.

Interestingly, the measuring time in biathlons represents another great challenge for EasyTime. Here, competitors ski on cross-country skiis and stop at certain places to shoot at targets with rifles carried by them. In order to measure time during biathlons, EasyTime needs to be modified slightly. In line with this, two measuring devices are need, and a special measuring device for counting hits. The first measuring device is dedicated to measuring the four laps of skiing, whilst the second is applied for counting the penalty laps. Each missed shot attracts one additional penalty lap. The measuring device for counting hits is described in EasyTime as a new agent. This agent is responsible for setting the number of additional penalty laps to be measured using the second measuring device. In contrast to the static initialization of the laps counter in EasyTime, a new request is demanded, i.e, a dynamic initialization of this laps counter needs to be implemented.

EasyTime could also be extended and used in some other application domains. For example, EasyTime could be employed as an electric shepherd for tracking livestock (cows, sheep, etc.) in the mountains. In this case, each animal would be labeled with a RFID tag that is controlled by crossing the measuring place twice a day. First, in the morning, when the animals go from their stalls and, second, in the evening, when they return to their stalls. Each crossing of the measuring place by the animal decrements a counter of herd-size for one. Essentially, the EasyTime tracking  system reports an error, when the counter is not decreased to zero within a specified time interval. In order for this tracking system to work properly, the herd-size counter has to be initialized twice a day (for example, at 12:00 am and 12:00 pm). Additionally, EasyTime could be used in the clothing industry for tracking cloth through the production. Clothing production consists of the following phases: preparing, sewing, ironing, adjusting, quality-control and packing~\cite{Fister:2008,Fister:2010}. The particular cloth origins during the preparation stage, where the parts of cutting patterns are collected into bundles, labeled with the RFID tags, and delivered for sewing. This transition of the bundle into the sewing room presents a starting point for the EasyTime tracking system. The other control points are, as follows: transition from sewing room into ironing, transition from ironing into adjusting, transition from adjusting into quality-control, and transition from quality-control into packing room that represents the finishing point of the cloth production. Note that these transitions act similarly to those transition areas in Ironman competitions. Usually, the cloth does not traverse through the production in any one-way because quality-control can return it to any of the past production phases. In this case, EasyTime could be used for tracking errors during clothing production.

\section{Conclusion}

The flexibility of the measuring system is a crucial objective in the development of universal software for measuring time in sporting
competitions. Therefore, the domain-specific language EasyTime was formally designed, which enables the quick adaptation of a measuring
system to the new requests of different sporting competitions. Preparing the measuring system for a new sporting competition with EasyTime requires the following: changing a program's source code that controls the processing of an agent, compiling a source code and restarting the agent. Using EasyTime in the real-world has shown that when measuring times in small sporting competitions, the organizers do not need to employ specialized and expensive companies any more. On the other hand, EasyTime can reduce the heavy configuration tasks of a measuring system for larger competitions, as well. In this paper, we explained how the formal semantics of EasyTime are mapped into LISA specifications from which a compiler is automatically generated. Despite the fact that mapping is not difficult, it is not trivial either, as some additional rules must be defined for attribute propagation. Moreover, we need to take care of error reporting (e.g., multiple declarations of variables).
In future work, EasyTime could be replaced by the domain-specific modeling language (DSML) \cite{Sprinkle:2009,Stuikys:2010,Vitiutinas:2011} that could additionally simplify the programming of a measuring system.

\bigskip{\small \smallskip\noindent Updated 9 June 2012.}
\end{document}